\documentclass[prb,twocolumn,aps,amssymb,amsmath,superscriptaddress,showpacs]{revtex4-1}

\usepackage{bm}
\usepackage{graphicx}
\usepackage{color}
\usepackage{mathtools}
\usepackage{amsfonts}
\usepackage{textcomp} 
\usepackage{microtype} 
\usepackage{epstopdf}
\usepackage{hyperref}
\usepackage[normalem]{ulem}
\hypersetup{colorlinks=true}
\usepackage[all]{hypcap} 
\graphicspath{{fig/}}

\begin{document}

\title{Topological phase transition in a two-species fermion system: Effects of a rotating trap potential or a synthetic gauge field}

\author{Shiuan-Fan Liou}
\affiliation{National High Magnetic Field Laboratory, Florida State University, Tallahassee, Florida 32310, USA}
\author{Zi-Xiang Hu}
\affiliation{Department of Physics, Chongqing University,Chongqing, 401331, People's Republic of China}
\author{Kun Yang}
\affiliation{National High Magnetic Field Laboratory, Florida State University, Tallahassee, Florida 32310, USA}

\begin{abstract}
We numerically investigate the quantum phases and phase transition in a system made of two species of fermionic atoms that interact with each other via $s$-wave Feshbach resonance, and are subject to rotation or a synthetic gauge field that puts the fermions at Landau level filling factor $\nu_f = 2$.  We show that the system undergoes a continuous quantum phase transition from a $\nu_f = 2$ fermionic integer quantum Hall state formed by atoms, to a $\nu_b = 1/2$ bosonic fractional quantum Hall state formed by bosonic diatomic molecules. In the disk geometry we use, these two different topological phases are distinguished by their different gapless edge excitation spectra, and the quantum phase transition between them is signaled by the closing of the energy gap in the bulk. Comparisons will be made with field theoretical predictions, and the case of $p$-wave pairing.
\end{abstract}

\date{\today}

\maketitle

\section{Introduction}

Topological phases of matter and the phase transitions between them have been the focus of much recent theoretical and experimental interests. The integer and fractional quantum Hall states, which are initially realized in a two-dimensional electron gas placed in strong magnetic fields, are prime examples of such topological phases. Trapped ultracold atoms constitute a unique experimental setup to study condensed-matter Hamiltonians in a clean and well-controlled environment~\cite{Cooper2008,RevModPhys.81.647}. One of the most interesting phenomena in the cold-atom system is the crossover from a weakly paired atomic fermionic superfluid to a strongly paired bosonic molecular superfluid as the pairing interaction is tuned through an $s$-wave Feshbach resonance~\cite{RevModPhys.80.1215}. When the trap potential of the cold atoms is rotating, the Coriolis force experienced by the atoms leads to an effective perpendicular magnetic field, and the quantum Hall states are expected in the fast rotation limit~\cite{Cooper2008,RevModPhys.81.647}. Recently, ways of engineering synthetic magnetic fields to realize quantum Hall states in cold-atom systems have been proposed, such as the strained optical lattice~\cite{PhysRevA.93.033640}, optical dressing~\cite{natureLin,PhysRevLett.93.033602} of atoms in a continuum, and laser-induced tunneling in an optical lattice~\cite{PhysRevLett.111.185301,PhysRevLett.111.185302, natphysicsKen}. Therefore, it is interesting to investigate what happens to these quantum Hall states in the presence of the pairing interaction between fermions.

With the $s$-wave pairing, it was pointed out by Yang and Zhai~\cite{PhysRevLett.100.030404} that in the quantum Hall regime, instead of a crossover, the system should undergo quantum phase transition(s) from a quantum Hall state formed by fermionic atoms at large positive detuning, to a topologically distinct quantum Hall state formed by bosonic molecules at large negative detuning. They used field-theoretical methods to study the special case in which the fermionic state is an integer quantum Hall state at Landau level filling factor $\nu_f = 2$, and showed that the system must undergo a quantum phase transition to a bosonic fractional quantum Hall state at $\nu_b = 1/2$ as a function of detuning, with the transition occurring near the Feshbach resonance (FR). These two phases, as well as a continuous quantum phase transition (QPT) between them, were indeed found in a numerical study of a Hubbard-like lattice model that includes only the fermionic atoms (or a single-channel model)~\cite{PhysRevB.96.161111} on a torus.

In the present work we perform a numerical study of the original two-channel model of Yang and Zhai~\cite{PhysRevLett.100.030404}, on a disk through the exact diagonalization method. There are three motivations to perform the present study. (i) It allows for a more direct and quantitative test of the predictions made by Yang and Zhai~\cite{PhysRevLett.100.030404}. (ii) The disk geometry is complementary to the torus geometry used by Ref.~\onlinecite{PhysRevB.96.161111}, as it allows for studies of the edge states, which are characteristics of the topological order. More importantly, it is directly relevant to experimental systems. (iii) In our previous study of the closely related system with $p$-wave pairing interaction between spinless fermions~\cite{PhysRevB.95.241106}, we found a new phase that is intermediate between the fermionic integer quantum Hall (FIQH) and bosonic fractional quantum Hall (BFQH) phases. Such a phase was missed by the effective-field theory~\cite{PhysRevLett.106.170403}. Therefore, in the $s$-wave pairing case, it is interesting to explore whether there is also a similar new phase.

The remainder of this paper is organized as follows. The microscopic model Hamiltonian we studied numerically is introduced in Sec. II, and the low-lying energy spectrum and ground state phase diagram are presented in Sec. III. In Sec. IV, we compare our present results with the $p$-wave pairing case and discuss the reason for the absence of the Bose-Fermi mixed phase in the $s$-wave case. Section V gives the summary and discussion.

\section{Model}

To study this QPT, we consider two species of fermions confined to a disk under rotation or a synthetic gauge field which gives the same effect as a strong magnetic field. We assume the Landau level spacing is so large that all particles stay in the lowest Landau level (LLL). We are interested in the case with the Landau level filling factor $\nu _{f} = 2$ (composed of two integer quantum Hall states, $\nu _{f \uparrow} = \nu _{f \downarrow} = 1$). When the system is tuned through $s$-wave FR, two fermions of different species can pair up to form an $s$-wave bosonic molecule with twice the ``charge." Between two fermions of the same kind, however, there is no interaction. Note that when all fermions pair up as bosons, the boson number is half of the total particle number, and the bosonic Landau level degeneracy, which is proportional to particle charge, is doubled. As a result, the bosonic filling factor $\nu _{b}$ will be $\frac{1}{4}$ of $\nu _{f}$, namely, $\nu _{b} = \frac{1}{2}$. In addition, the corresponding bosonic magnetic length square $l _{b} ^{2}$, which is inversely proportional to particle charge, will be half of the fermionic magnetic length square $l _{f} ^{2}$, $l _{b} ^{2} = \frac{1}{2} l _{f} ^{2}$. With rotational symmetry, this system is described by the following Hamiltonian
\begin{eqnarray}\label{hamiltonian}
H &=& \delta  \sum\limits _{m} ( b _{m} ^{\dagger}  b _{m}  -  \sum\limits _{\sigma = \uparrow, \downarrow } f _{\sigma \, m} ^{\dagger} f _{\sigma m} ) \nonumber \\
&+& \sum\limits _{m _{1}, m _{2}, m _{3}}  (g _{m _{1}, m _{2}, m _{3}} b _{m _{1}} ^{\dagger} f _{\uparrow, m _{2}} f _{\downarrow, m _{3}}  +  \mbox{H.c.} ) \nonumber \\
&+&  \sum\limits _{m _{1}, m _{2}, m _{3}, m _{4}}  v ^{(0)} _{m _{1}, m _{2}, m _{3}, m _{4}}  b _{m _{1}} ^{\dagger}  b _{m _{2}} ^{\dagger} b _{m _{3}}  b _{m _{4}},
\end{eqnarray}
where $m$ is the angular momentum of the single-particle orbital in the LLL with $b _{m} (b _{m} ^{\dagger})$ and $f _{\sigma , m} (f _{\sigma , m} ^{\dagger})$ being the corresponding annihilation (creation) operators for bosons and fermions.

The first term (chemical potential) controls whether atoms should stay unbound or form molecules. $\delta$ in this term is the detuning referring to the energy difference between unbound and paired fermions. The ``detuning'' we used here is from FR. The second term describes the pairing interaction through $s$-wave FR. The matrix element of this term can be written as $g _{m _{1}, m _{2}, m _{3}}=g \delta _{m _{1}, M} \, \delta _{M, m _{2} + m _{3}} \, \langle 0, M| m_{2}, m _{3} \rangle$, where $g$ represents the strength of the $s$-wave pairing, $| m _{2}, m _{3} \rangle \equiv |m _{2} \rangle _{\uparrow} \otimes | m _{3} \rangle _{\downarrow}$ is a two-body state with two fermions of different species with angular momenta $m _{2}$ and $m _{3}$, and $|0, M \rangle$ is a two-body state with the relative angular momentum equal to zero and center-of-mass angular momentum $M$. This term allows only two fermions of different species with relative angular momentum $\Delta m = 0$ to pair up, and the formed bosonic molecule will have angular momentum $m _{1} = M = m _{2} + m _{3}$ based on angular momentum conservation; in other words, the boson itself has no intrinsic angular momentum ($s$ wave). The Clebsch-Gordan-like coefficient $\langle 0 , M| m_{2}, m _{3} \rangle$ can be evaluated through\\
\begin{equation}
\begin{aligned}
&\left\langle \Delta m , M | m _{1}, m _{2} \right\rangle \equiv \\
& \times \frac{1}{\sqrt{(2 \pi) ^{4}  2 ^{( \Delta m + M + m _{1} + m _{2} )}  \Delta m!  M!  m _{1}! m _{2}!  l ^{8}}} \\
&\int d ^{2} z_{1} \int d ^{2} z _{2} ( \frac{z _{1} ^{\ast} - z _{2} ^{\ast}}{\sqrt{2} l} ) ^{\Delta m} (\frac{z _{1} ^{\ast} + z _{2} ^{\ast} }{\sqrt{2}
l}) ^{M} ( \frac{z _{1}}{l} ) ^{m _{1}} ( \frac{z _{2}}{l} ) ^{m _{2}} \\
& \hspace*{6cm} \times e^{- \frac{|z _{1}| ^{2} + |z _{2}| ^{2}}{2 l ^{2}}}  \\
& = \, \sqrt{\frac{\Delta m! \, M!}{\left( 2 \pi \right) ^{4}  2 ^{\Delta m + M} \,  m _{1}!  m _{2}!}} \\
& \times \sum\limits_{n} \, (-1) ^{\Delta m - n} \, C ^{n} _{m _{1}} \, C ^{\Delta m - n} _{m _{2}} \,  \delta _{\Delta m + M, m _{1} + m _{2}},
\end{aligned}
\end{equation}
where $z _{a} \equiv x _{a} + i y _{a}$ is the complex coordinate of the $a$th particle on a disk in the LLL, $d ^{2} z _{a} = dx _{a} dy _{a}$, $\Delta m$ is the relative angular momentum of a pair, and $C ^{n} _{m} = \frac{m!}{n! \left( m - n \right)!}$ is the binomial coefficient. The sum of $n$ is over all integers bounded at max$\left(0, \Delta m - m _{2} \right) \leq n \leq$ min$\left(\Delta m, m _{1} \right)$.
The last term in Eq.~(\ref{hamiltonian}) is the two-body repulsive interaction between bosons for stabilizing the BFQH state with $\nu _{b} = \frac{1}{2}$.
Here we include only the $zero$th Haldane pseudopotential~\cite{PhysRevLett.51.605}, which makes the $\nu_b=1/2$ Laughlin wave function the exact ground state when we have only bosonic molecules. The matrix element is expressed as $v ^{\left( 0 \right)} _{m _{1}, m _{2}, m _{3}, m _{4}} = v ^{\left( 0 \right)} \sum _{M} \langle m _{1} , m _{2} | 0, M \rangle \langle 0, M | m _{3} , m _{4} \rangle$, where $v ^{\left( 0 \right)}$ denotes the strength of the $zero$-th order Haldane pseudopotential. In our system, we are considering the case at temperature $T$ = 0 in which changing $v ^{\left( 0 \right)}$ has no other effect than changing the overall energy scale. In reality, when $T > 0$, this statement is still valid as long as $v ^{\left( 0 \right)}$ is much larger than $k _{B} T$. Thus, we choose $v ^{\left( 0 \right)} = 1$ in our calculation.

In our model, both the total charge $N _{tot}$ and the total angular momentum $M _{tot}$ are good quantum numbers. $N _{tot}$ is the sum of the numbers of two species of fermions and twice the number of bosons:
\begin{equation}
N _{tot} =  2 N _{b} + N _{f \uparrow} + N _{f \downarrow} = \sum\limits _{m} ( 2 b _{m} ^{\dagger} b _{m} +  \sum\limits _{\sigma = \uparrow , \downarrow} f _{\sigma m} ^{\dagger} f _{\sigma m}  ).
\end{equation}
\noindent The prefactor 2 in the bosonic part comes from the fact that a boson consists of two fermions. The total angular momentum $M _{tot}$ is the sum over all orbitals occupied by bosons and fermions:
\begin{equation}
M _{tot} = \sum _{m} \, m ( b _{m} ^{\dagger} b _{m} + \sum _{\sigma} f _{m \sigma} ^{\dagger} f _{m \sigma} ).
\end{equation}
In our numerical calculation on disk geometry, we use $N _{tot}$ and $M _{tot}$ to label the sector where the calculations are performed.

\section{Numerical results}

The Hamiltonian $H$ in Eq.~(\ref{hamiltonian}) has two limits. When $\delta > 0$ and $|\delta| \gg g$, unpaired fermions have lower energy than bosons, so the chemical potential term drives the ground state of the system toward being a FIQH state at $\nu _{f} = 2$ (composed of two copies of FIQH state $\nu _{f \sigma} = 1$), which does not need to be stabilized by any interaction between the fermions. On the other hand, for $\delta < 0$ and $|\delta| \gg g$, bosonic molecules have lower energy and dominate in the ground state of the system. Owing to the existence of the bosonic repulsive two-body interaction, the ground state becomes a Laughlin-type BFQH state with $\nu _{b} = \frac{1}{2}$. In the following, we will inspect the low-energy spectra to distinguish between these two phases.

\begin{table}
\begin{tabular}{ |c | p{1.5cm} | p{1.5cm} | p{1.5cm} | p{1.7cm}| }
    \hline
    $\Delta M$ & 0 & 1 & 2 & 3 \\ \hline
    $U \left( 1 \right)$ & $\epsilon _{0}$ & $\epsilon _{0} + \epsilon _{1}$ & $\epsilon _{0} + 2 \epsilon _{1}$ \, $\epsilon _{0} + \epsilon _{2}$ & $\epsilon _{0} + 3 \epsilon _{1}$ \, \, $\epsilon _{0} + \epsilon _{2} + \epsilon _{1}$  $\epsilon _{0} + \epsilon _{3}$  \\ \hline
    $U \left( 1 \right) \times U \left( 1 \right)$ & $\epsilon _{0}$ & $\epsilon _{0} + \epsilon _{1} ^{\alpha}$ & $\epsilon _{0} + 2 \epsilon _{1} ^{\alpha}$  $\epsilon _{0} + \epsilon _{2} ^{\alpha}$ \, $\epsilon _{0} + \epsilon _{1} ^{\uparrow} + \epsilon _{1} ^{\downarrow}$ & $\epsilon _{0} + 3 \epsilon _{1} ^{\alpha}$ \, \, $\epsilon _{0} + \epsilon _{2} ^{\alpha} + \epsilon _{1} ^{\alpha}$ \,\,  $\epsilon _{0} + \epsilon _{3} ^{\alpha}$ \, \, \, $\epsilon _{0} + \epsilon _{2} ^{\uparrow} + \epsilon _{1} ^{\downarrow}$ \,\, $\epsilon _{0} + \epsilon _{1} ^{\uparrow} + \epsilon _{2} ^{\downarrow}$ \, $\epsilon _{0} + 2 \epsilon _{1} ^{\uparrow} + \epsilon _{1} ^{\downarrow}$ \, $\epsilon _{0} + 2 \epsilon _{1} ^{\downarrow} + \epsilon _{1} ^{\uparrow}$ \\ \hline
\end{tabular}
\caption{Edge-state counting. $\Delta M$ is the exceeded angular momentum compared to $M _{gs}$; $\epsilon _{0}$ is the root configuration of the Laughlin state, and $\epsilon _{i} ^{\alpha}$ represents the change in the configuration where one $\alpha$-type particle changes its angular momentum by $i$, where $\alpha = \uparrow$ or $\downarrow.$ In our model, the edge counting for fermionic Hilbert space corresponds to the $U \left( 1 \right) \times U \left( 1 \right)$ case, and that for bosonic Hilbert space corresponds to the $U \left( 1 \right)$ case. Note that some possibilities may be prohibited by the number of extra orbitals.}
\label{edg_cnt_tb}
\end{table}

\begin{figure}[!t]\label{edg_cnt}
\centerline{\includegraphics[width=9cm]{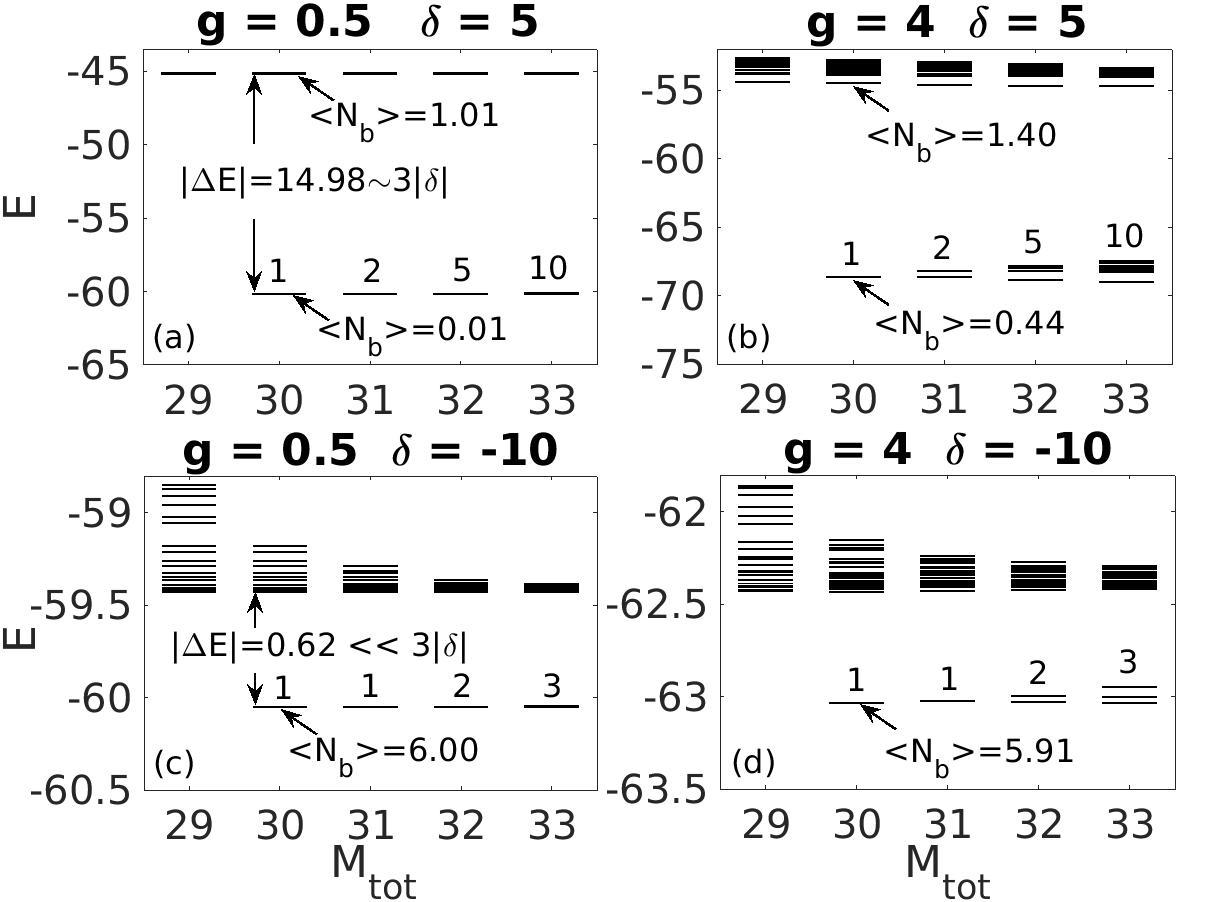}}
\caption{The energy spectra for a system on a disk with $N_{tot}=12$ fermions, given 9 fermionic orbitals and 18 bosonic orbitals. The 20 lowest energy states are plotted for each $M
_{tot}$. The ground state at $M_{tot}=M_{gs}=30$ is separated by a large gap from all excited states for (a) $g$ = 0.5 and $\delta = 5$ and (b) $g$ = 4 and $\delta = 5$, which are
under the FIQH limit. 
The numbers right above the low-lying states represent their (near) degeneracy. At $M _{tot} > M _{gs}$, many low-lying states are found and their numbers are consistent with the edge state counting. We find a different set of low-lying states for $M _{tot}\ge M_{gs}=30$ for (c) $g$ = 0.5 and $\delta = -10$ and (d) $g$ = 4 and $\delta = -10$, which are under the BFQH limit. These correspond to the Laughlin-like ground state at $M_{tot}=M_{gs}=30$ and edge states for $M _{tot} > M_{gs}=30$. The numbers of these states indicated in the plots match the expected numbers of edge states. $\left\langle N _{b} \right\rangle$ indicate the expectation values of boson numbers in the corresponding states pointed to by small arrows. The Hilbert space dimensions for the situation where $N _{tot} / 2$ are bosons and $N _{tot} / 2$ are fermions, for fermions and bosons at $M _{tot} = 30$ are about $\left( \frac{6 ! !}{3! \, 3!} \right) ^{2} = 400$ and $\frac{14!}{11! \, 3!} = 364$, respectively. }
\end{figure}

Figure~\ref{edg_cnt} shows the low energy spectra for a system with $N _{tot} = 12$ given 9 up (down) fermionic orbitals and 18 bosonic orbitals at $M _{tot} = 29$ to $33$. Instead of providing the least orbital numbers $\left( \frac{ \left( N _{tot}/2 \, - 1 \right)}{\nu _{b \left( f \sigma \right)}} + 1 \right)$, we give three more orbitals for each species of fermions and seven more orbitals for bosons to allow the appearance of the edge states~\cite{wen:IJMPB}. Under the FIQH limit, with the least orbitals the system will have the lowest-energy state only when it forms a FIQH state in which there are no bosons and all fermionic orbitals are occupied, namely, at $M _{tot} = M _{gs}$, with
\begin{equation}
M _{gs} = \frac{N _{tot}}{2} ( \frac{N _{tot}}{2} - 1).
\end{equation}
For this case with $N _{tot} = 12$, $M _{gs} = 30$. With extra orbitals, edge states degenerate with the FIQH state are expected to appear at $M _{tot} > M _{gs}$ but not at $M _{tot} < M _{gs}$. Since we have two species of fermions ($\uparrow$ and $\downarrow$), the Hilbert space of the fermionic part of the system is the tensor product of the Hilbert space of each species of fermion. As a result, the counting of the fermionic edge states is $U \left( 1 \right) \times U \left( 1 \right)$. Some examples are shown in Table~\ref{edg_cnt_tb}~\cite{wen:IJMPB}. By counting and comparing the low-lying states with the numbers of edge states at various $M _{tot}$, we can demonstrate the system forms a FIQH state. In Figs.~\ref{edg_cnt}(a) and \ref{edg_cnt}(b) with $\delta = 5$ at $g = 0.5$ (weak coupling) and $g = 4$ (strong coupling), the consistency of the numbers of the low-lying states (the numbers right above the low-lying states) and the edge states illustrates that the system stays in the FIQH phase. Note that the low-lying states are no longer exactly degenerate due to the existence of the pairing interaction. Moreover, we also inspect the boson numbers in the ground and the first excited state at $M _{gs}$. The former is very close to zero, and the latter is about 1, as expected. The energy gap between the ground state and the first excited state $\Delta E$, mainly contributed by the chemical-potential term, which is about 3$|\delta|$ (losing two fermions and gaining a boson), is also observed in Fig.~\ref{edg_cnt}(a), as expected.

In the BFQH limit $\delta\rightarrow -\infty$, the system contains $N _{b}= N _{tot} / 2$ bosonic molecules at $\nu = \frac{1}{2}$, and has the Laughlin wave function
\begin{equation}
\psi _{\nu = \frac{1}{2}} \left( z _{1}, z _{2} \cdots \right) \, = \, \left[ \prod _{i < j} ^{N_b} \, \left( z _{i} - z _{j} \right) ^{2} \right] \, e ^{-\sum\limits _{k
= 1}^{N _{b}} \, \frac{|z _{k}| ^{2}}{4 \, l ^{2}}}
\end{equation}
as its exact ground state. The corresponding total angular momentum is $M _{tot} = 2 \times N _{b} \left( N _{b} - 1 \right) / 2 = M _{gs}$, the same as that in the FIQH limit. To identify the BFQH phase, we compare the numbers of the low-lying states with the numbers of edge states of Laughlin-type states which correspond to $U \left( 1 \right)$ counting in Table~\ref{edg_cnt_tb}. In Figs.~\ref{edg_cnt}(c) and \ref{edg_cnt}(d), we see that both countings are consistent meaning that the system forms a Laughlin-type BFQH state with $\nu = \frac{1}{2}$, at $\delta=-10$. Besides, the low-lying states at $M _{gs}$ in both weak- and strong-coupling regimes have nearly six bosons, the maximum number of bosons. In Figs.~\ref{edg_cnt}(b) and \ref{edg_cnt}(d), the bigger deviation of the particle numbers from the expected values in the strong-coupling regime tells us that the pairing term provides fermions more chances (bigger matrix elements) to jump back and forth between bound and unbound states. One thing very different from the FIQH limit is that the energy gap at $M _{gs}$ in Fig.~\ref{edg_cnt}(c) is much smaller than 3$|\delta|$ because the low-energy excited states are still made dominantly of bosons (almost no fermions) with different angular momentum configurations. It turns out the bosonic two-body interaction becomes the main contributor to the energy gap rather than the chemical-potential term in this case.

\begin{figure}[!t]\label{d_gap}
\centerline{\includegraphics[width=8.5cm]{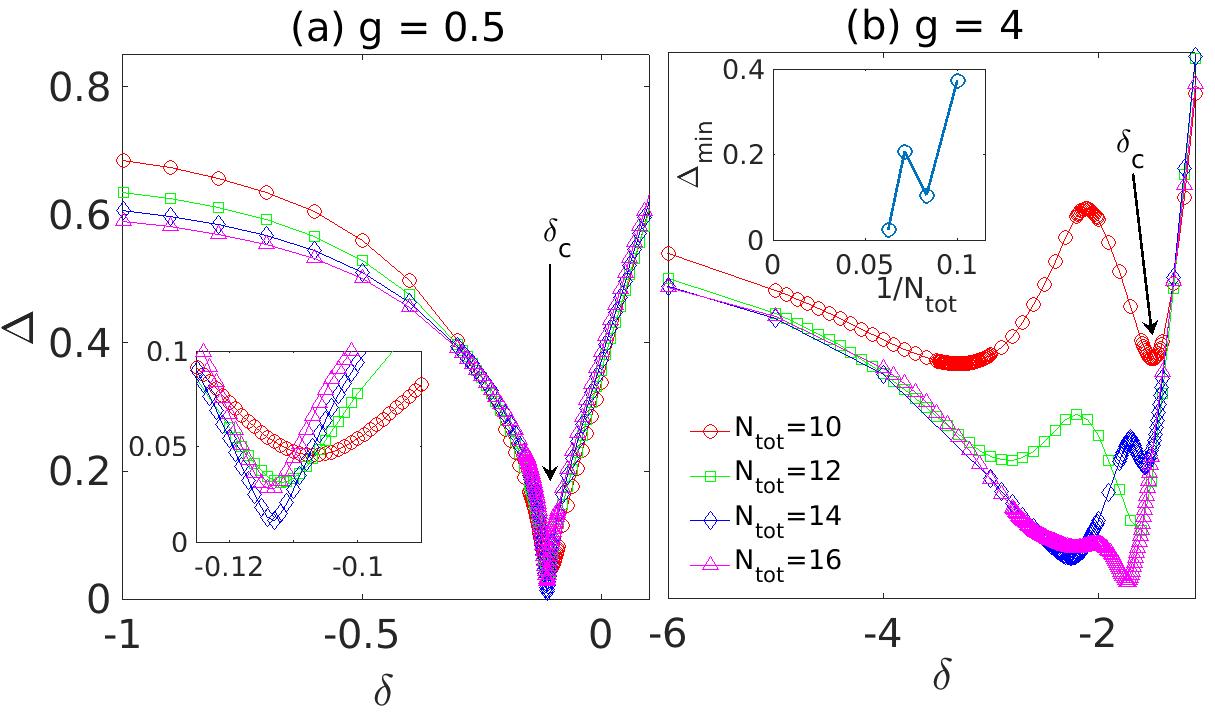}}
\caption{(Color online) Plot of the energy gap ($\Delta \equiv E _{1} - E _{0}$, where $E _{0}$ and $E _{1}$ are the energies of the ground state and the first excited state in the $M _{gs}$ sector) versus $\delta$ for systems with $N _{tot} = 10, 12, 14$, and 16 at $M _{gs}$. (a) For $g$ = 0.5. There is one gap closing point $\delta _{c}$ for each curve.
Inset: blowup of the gap-closing region. (b) For $g$ = 4. There is still one gap-closing point for each curve, despite the existence of another local gap minimum on the left which fades away as system size increases. Inset: values of gap minimums corresponding to the phase boundary versus inverse of system size.}
\end{figure}

\begin{figure}[!t]\label{nb_mu}
\centerline{\includegraphics[width=8cm]{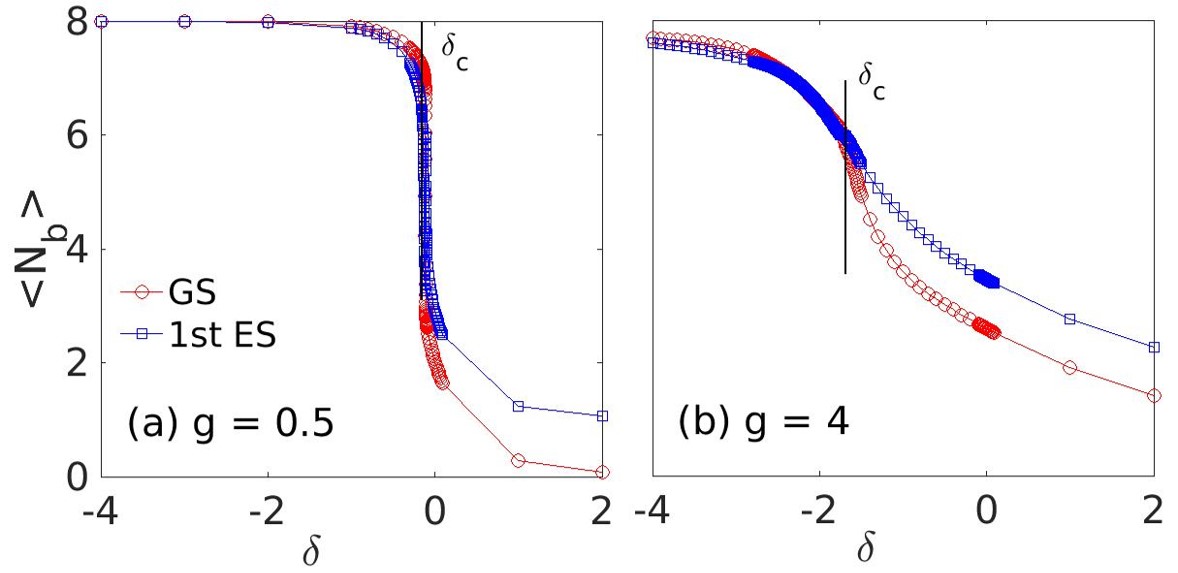}}
\caption{(Color online)  The expectation values of boson numbers $\langle N _{b} \rangle$ in the ground state and the first excited state versus $\delta$ for the system with $N _{tot} = 16$ at (a) $g = 0.5$ and (b) $g = 4$. The vertical black lines indicate the critical points.}
\end{figure}

In order to reach the appropriate ground states for continuously varying $\delta$ and to save calculation efforts, given a specific $N _{tot}$, from now on we give the fewest orbitals $\left[\frac{\left( N _{tot} / 2 \, - 1 \right)}{\nu _{b \left( f \sigma \right)}} + 1 \right]$ for each kind of particle and focus on the $M _{gs}$ sector, which is the same for any $\delta$ in our calculation. To explore the phase diagram, we drive the system from the FIQH phase to the BFQH phase by changing $\delta$ at various $g$. Since topological quantum phase transitions between distinct gapped phases must be associated with gap closing, we investigate the behavior of the gap $\Delta$ while varying $\delta$ in Figs.~\ref{d_gap}(a) and \ref{d_gap}(b) at $g = 0.5$ and $g = 4$, with $\Delta$ defined as the energy difference between the first excited state and the ground state. Four system sizes with $N _{tot} = 10, 12, 14$, and $16$ are considered. In Figs.~\ref{d_gap}(a) (weak-coupling regime) and \ref{d_gap}(b) (strong-coupling regime), the observation of one gap-closing point signifies there exists only one phase boundary, separating the FIQH and BFQH phases, in the entire phase diagram. Although another local gap minimum in the strong-coupling regime in Fig.~\ref{d_gap}(b) is observed at small sizes, it fades away as the system size grows and does not signal gap closing or another phase boundary. From Fig.~\ref{d_gap}[see also the inset in Fig.~\ref{d_gap}(a)], the nonzero gap (close to zero but not exactly zero) at $\delta _{c}$ is due to the finite-size effect, and it approaches zero with increasing system size, consistent with the feature of a continuous phase transition. The inset of Fig.~\ref{d_gap}(b) shows the values of the gap minimums corresponding to the phase boundary decrease with system size, although the values oscillate with the parity (even or odd) of the particle number, which is also due to the finite-size effect. In addition, we inspect the average boson numbers $\langle N _{b} \rangle$ of the ground state and the first excited state in Figs.~\ref{nb_mu}(a) and \ref{nb_mu}(b) and find $\langle N _{b} \rangle$ of the ground states in both pairing regimes increase smoothly and monotonically when $\delta$ changes from positive (FIQH phase) to negative (BFQH phase). This is also strong evidence of a continuous phase transition.

\begin{figure}[!ht]\label{phs_dig}
\centerline{\includegraphics[width=8cm]{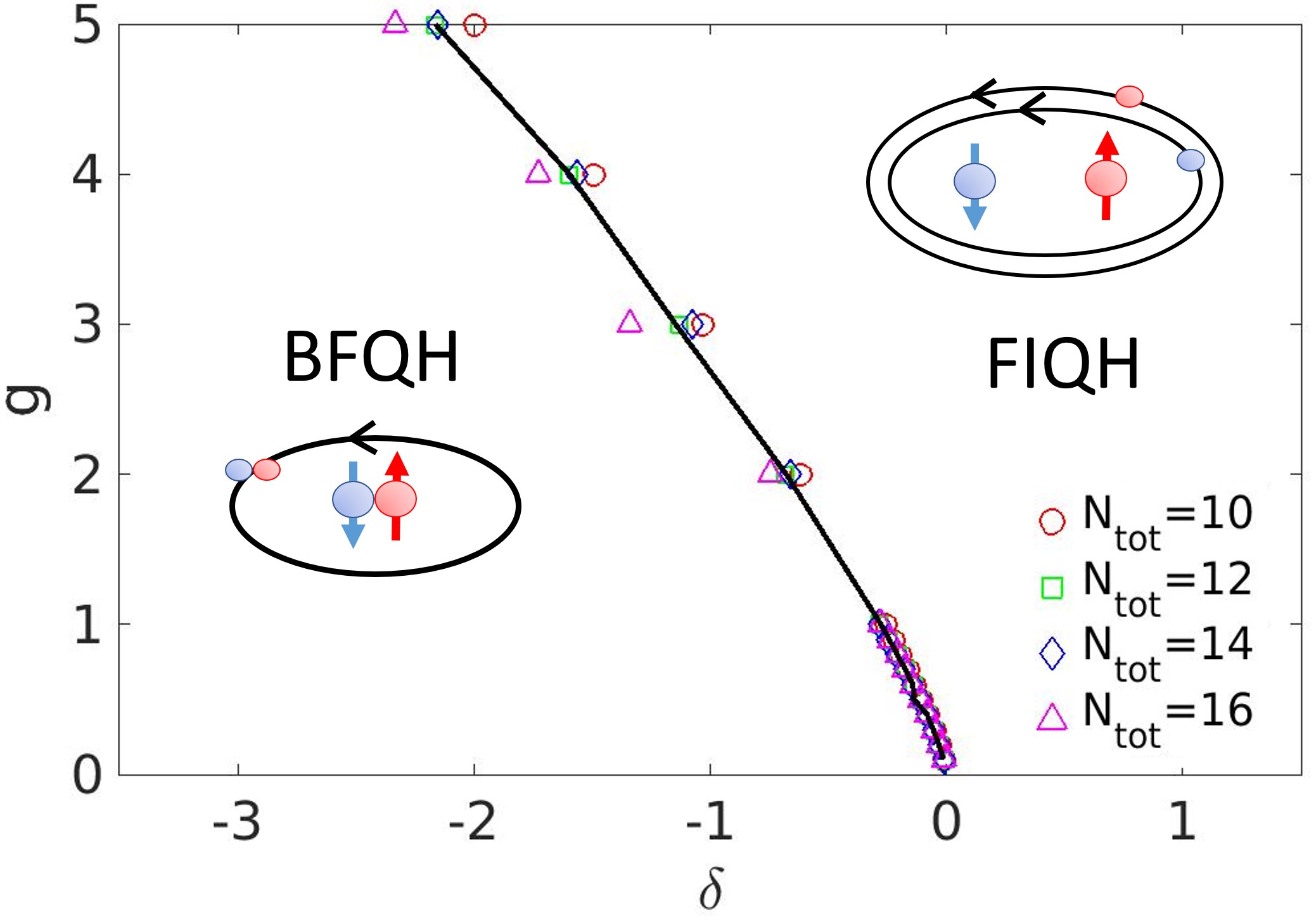}}
\caption{(Color online) Phase diagram with system sizes $N _{tot} = 10, 12, 14$, and $16$. The FIQH state is characterized by two branches of edge modes, and the BFQH state has only one branch.}
\end{figure}

\begin{figure}[!t]\label{nf_ratio}
\centerline{\includegraphics[width=8cm]{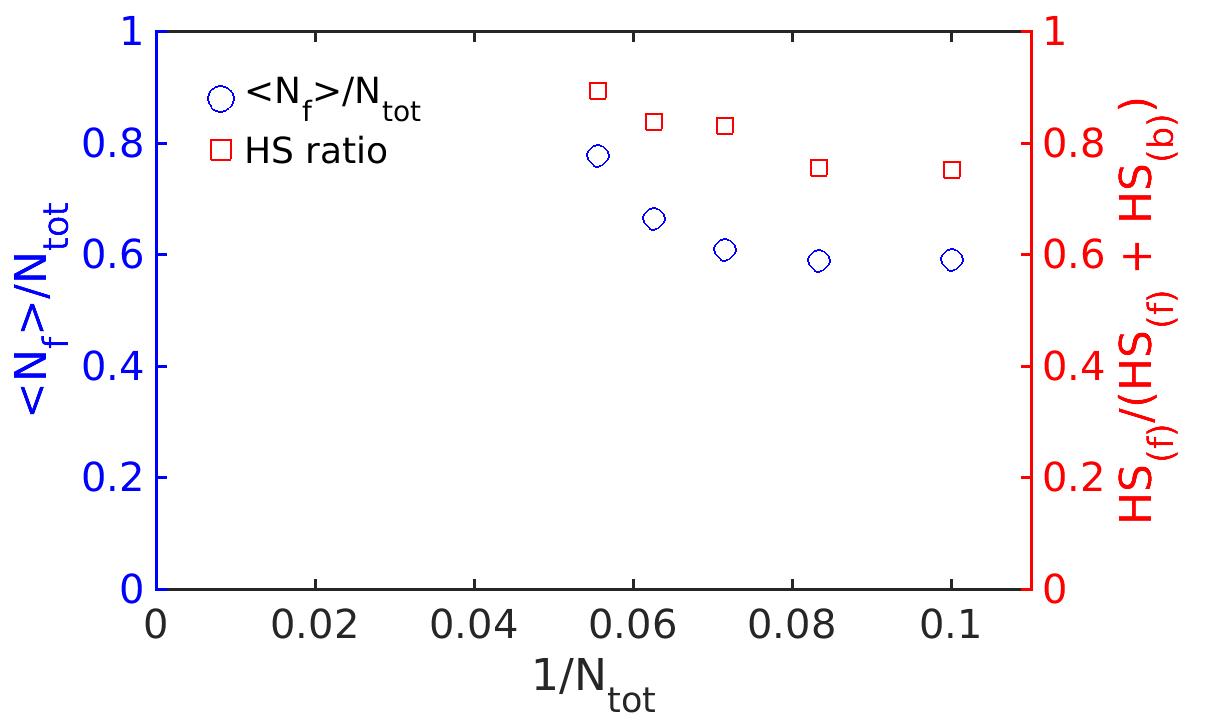}}
\caption{(Color online) The average fermion fraction $\frac{<N_{f}>}{N_{tot}}$ in the ground state at $g = 1$, $\delta = 0$, and $v ^{\left( 0 \right)} = 0$ versus the system size on the left axis and the ratio of the size of the fermionic Hilbert space $HS_{(f)}$ to the sum of the sizes of the fermionic and bosonic Hilbert spaces $HS_{(f)} + HS_{(b)}$ in the spin-1/2 case vs the system size on the right axis, regarding $HS_{(f)}$ and $HS_{(b)}$ in spinless case as a unit. Five system sizes are explored with $N _{tot} = 10, 12, 14, 16$, and $18$.}
\end{figure}

Based on the locations of the gap closings, we obtain a phase diagram in Fig.~\ref{phs_dig}. This phase diagram possesses a single phase boundary, which starts at zero detuning ($\delta=0$) in the weak-coupling (or narrow-resonance) limit ($g=0$) and moves toward negative detuning ($\delta < 0$) with increasing $g$. These features of the phase diagram are in qualitative agreement with the predictions of Yang and Zhai~\citep{PhysRevLett.100.030404}.

\section{Comparison with $p$-wave pairing}
In the $p$-wave pairing case (spinless case) in our previous work\citep{PhysRevB.95.241106}, we found an intermediate phase called the coherent Bose-Fermi mixture phase in the strong-coupling region between the FIQH phase at $\nu _{f} = 1$ and the BFQH phase at $\nu _{b} = 1/4$. However, we do not find such a phase in our present study (spin-1/2 case). The purpose of this section is to analyze the origin of this difference.
To get a hint, in Fig.~\ref{nf_ratio}, where the $g$ term dominates ($g = 1$, $\delta = 0$ and $v ^{(0)} = 0$), we find that the fermion fraction of the ground state increases with the system size, different from the observation of a constant fermion fraction (independent of system size) in the spinless case. This suggests the fermions prefer to stay unbound instead of forming bound molecules, despite the fact that the Hamiltonian is purely off-diagonal between bound and unbound states, thus favoring an equal-weight mixture between the two.
The increase in fermion number, therefore, must be due to the larger phase space available to fermions compared to bosons. This suggests that as system size increases, the phase-space sizes grow at an unequal rate for fermions and bosons in a way that favors the former.

To quantify this idea, we estimate the size of the fermionic Hilbert space for each species of fermions, which is $\frac{N _{orb} ^{f}!}{N_{f}! \left( N _{orb} ^{f} - N _{f} \right)!}$ (choosing $N _{f}$ positions for fermions from $N _{orb} ^{f}$ orbitals), where $N _{f}$ is the fermion number and $N _{orb} ^{f}$ is the fermionic orbital number. For the spinless case which has $N _{tot}$ fermionic orbitals and $N _{tot} / 2$ fermions in the special situation of an equal mixture of bosons and fermions, the size of the Hilbert space is $\frac{N _{tot}!}{\left( N _{tot} / 2 \right)! \left( N _{tot} / 2 \right)!}$; for the spin-1/2 case in which each species of fermions has $N _{tot} / 2$ fermionic orbitals and $N _{tot} / 4$ particles, the size is about $\left[ \frac{\left( N _{tot} / 2 \right)!}{\left( N _{tot} / 4 \right)! \left( N _{tot} / 4 \right)!} \right] ^{2}$, where the power of 2 originates from the two species. On the other hand, the size of the bosonic Hilbert space can be evaluated as $\frac{\left( N _{orb} ^{b} + N _{b} - 1 \right)!}{\left( N _{orb} ^{b} - 1 \right)! N _{b}!}$, with $N _{b}$ being the boson number and $N _{orb} ^{b}$ being the bosonic orbital number, based on the nature of bosons in which more than two bosons can occupy the same orbitals. The spinless case with two times as many bosonic orbitals has the Hilbert space size $\frac{\left( 2 N _{tot} + N _{tot}/4 \; - \; 1 \right)!}{\left( 2 N _{tot} - 1 \right)! \left( N _{tot} /4 \right)!}$, and the size in the spin-1/2 case is $\frac{\left( N _{tot} + N _{tot}/4 \; - \; 1 \right)!}{\left( N _{tot} - 1 \right)! \left( N _{tot} /4 \right)!}$.

When the system size is small, both Hilbert spaces of fermionic and bosonic parts in the spin-1/2 case are slightly smaller than but comparable with their corresponding Hilbert spaces in the spinless case.
When the system size increases ($N _{tot} \rightarrow \infty$ limit), the fermionic Hilbert space in both cases has the same expansion rate as system size ($\sim e ^{0.301 * N _{tot}}$), but the bosonic Hilbert space in the spin-1/2 case has a small expansion rate ($\sim e ^{0.271 * N _{tot}}$) compared to the rate ($\sim e ^{0.341 * N _{tot}}$) in the spinless case due to the fact that given the same number of bosons $N_{b}$, the spinless case at $\nu _{b} = 1/4$ will have $4 N_{b}$ bosonic orbitals, two times more than that in the spin-1/2 case at $\nu _{b} = 1/2$, leading to significantly different Hilbert space sizes in these two cases, especially with a large number of bosons. Therefore, in the thermodynamic limit, the Hilbert space of the fermionic and bosonic parts in the spin-1/2 case are no longer compatible, resulting in the dominance of the fermions, as shown in Fig.~\ref{nf_ratio}. In Fig.~\ref{nf_ratio}, the ratio of the size of the fermionic Hilbert space $HS_{(f)}$ to the sum of the sizes of the fermionic and bosonic Hilbert spaces $HS_{(f)} + HS _{(b)}$ in the spin-1/2 case, regarding the corresponding sizes in the spinless case as a unit, increases with increasing system size. Its trend is consistent with that of the fermion fraction, indicating the significance of the small expansion rate in the bosonic Hilbert space (compared to the spinless case) even though the difference is only the coefficient on the exponent. This observation also explains the appearance of two gap minima in Fig.~\ref{d_gap}(b). For a small system with $N _{tot} = 10$,
the system is trying to form a coherent Bose-Fermi mixture phase similar to that in our previous work~\cite{PhysRevB.95.241106} and two apparent phase boundaries that correspond to the two gap minima. As the system size increases, this intermediate phase is disfavored, and the second gap minima disappears accordingly.

Moreover, comparing the possible combinations of fermions to form molecules, the spin-1/2 case has $N _{tot} ^{2} / 4$ ways to pair between up- and down-spin fermions, about a factor of 2 less than the spinless case, which has $N _{tot} ( N _{tot} - 1)/2$ ways to pair. This means that the spin-1/2 case has fewer channels for fermions to resonate between bound and unbound states. This also disfavors the coherent Bose-Fermi mixture phase in the present case.

\section{Summary and discussion}

Using exact diagonalization of finite-size systems on a disk, we investigated the topological phase transition from a fermionic integer quantum Hall state composed of two copies of an integer quantum Hall state of two species of fermions to a bosonic fractional quantum Hall state made of bosonic molecules, driven by an $s$-wave Feshbach resonance. We demonstrated the existence of a continuous phase transition from the fermionic integer quantum Hall phase to the bosonic fractional quantum Hall phase and provided a phase diagram which contains a single phase boundary. Our results agree with earlier theoretical predictions\cite{PhysRevLett.100.030404}, and a recent numerical work\cite{PhysRevB.96.161111} based on a different model. In addition, we argued that the absence of the coherent Bose-Fermi mixed phase found in a related work\cite{PhysRevB.95.241106} is due to the imbalance between bosonic and fermionic Hilbert space sizes in the present case.

\acknowledgements
S.-F.L. and K.Y. are supported by DOE Grant No. DE-SC0002140. Part of this work is performed at the National High Magnetic Field Laboratory, which is supported by National Science Foundation Cooperative Agreements No. DMR-1157490 and No. DMR-1644779, and the State of Florida. Z.-X.H. is supported by NSFC under Projects
No. 11674041, and No. 91630205 and the Chongqing Research Program of Basic Research
and Frontier Technology Grant No. cstc2017jcyjAX0084.

\bibliography{ref}

\end{document}